\documentstyle[12pt]{article}
\begin{document}
\setlength{\unitlength}{1mm}
{\hfill  JINR E2-94-185, 1994 } \vspace*{2cm} \\
\begin{center}
{\Large\bf On Higher Derivative Gravity In Two Dimensions}
\end{center}
\begin{center}
{\large\bf  S.N.Solodukhin$^{\ast}$}
\end{center}
\begin{center}
{\bf Bogoliubov Laboratory of Theoretical Physics, Joint Institute for
Nuclear Research, Head Post Office, P.O.Box 79, Moscow, Russia}
\end{center}
\vspace*{2cm}
\abstract
The 2D gravity described by the action which is an arbitrary
function of the scalar curvature $f(R)$ is considered. The classical vacuum
solutions
are analyzed. The one-loop renormalizability is studied. For the function
$f=R \ln R$ the model coupled with scalar (conformal) matter is
exactly integrated and is shown to describe a black hole of the 2D dilaton
gravity
type. The influence of one-loop quantum corrections on a classical
black hole configuration is studied by  including
the Liouville-Polyakov term. The resulting model turns out to be exactly
solvable. The general solution is analyzed and shown to be free from the
space-time
singularities for a certain number of scalar fields.
\begin{center}
{\it PACS number(s): 04.60.+n, 12.25.+e}
\end{center}
\vskip 4cm
\noindent $^{ \ast}$ e-mail: solodukhin@main1.jinr.dubna.su
\newpage
\section{Introduction}
\setcounter{equation}0

Recently, much attention has been paid to the investigation of models of
two-dimensional (2D) gravity. It is well known that the Einstein-Hilbert action
in
two dimensions coincides with the topological Euler number and, therefore, does
not
determine any dynamics for gravitational (metrical) degrees of freedom. Hence,
one should
consider some alternative dynamical descriptions of 2D gravity. One of the
simplest
models, mainly inspired by the string theory, is dilaton gravity [1],
gravitational
variables are the dilaton and metric fields $(\phi, g_{\mu\nu})$. In empty
(without matter) space the classical equations of motion are exactly integrated
[1] and the solution describes the 2D black hole. On the quantum level, it has
been
shown that this model is renormalizable [2]. The coupling with conformal matter
is again exactly solvable classically and the solutions are configurations
describing the formation of a black hole by collapsing matter [3].

An other way is to formulate the theory of 2D gravity in the framework of
a consistent gauge approach. Independent variables are now vielbeins and
the Lorentz connection $(e^a, \omega^a_{\ b})$.  The theory with Lagrangian
quadratic in curvature $R$ and torsion $T$ [4] was shown to be exactly
solvable [5]. One class of the solutions contains the de Sitter space-time with
zero torsion. Other solutions are of the black hole type [5]. Generally, one
can consider the Lagrangian to be an arbitrary (not-necessarily quadratic)
function of curvature and torsion [6]. Such a theory has essentially the same
type of classical solutions.

Describing the gravitational degrees of freedom on the 2D manifold $M^2$ only
by
 the metric ($g_{\mu\nu})$ without introducing any additional variables,
one considers the following action:
\begin{equation}
S=\int\limits_{M^2}^{}d^2 z \sqrt{-g} f(R),
\end{equation}
where $f(R)$ is, in principle, arbitrary (non-linear) function of the scalar
curvature $R$ determined with respect to the 2D metric $g_{\mu\nu}$.
Theories of such type were studied in higher dimensions [7] and
in two dimensions [8,9]. Was observed that the theory (1.1) is equivalent to
some type of scalar-tensor $(\phi, g_{\mu\nu})$ theory of gravity. Moreover,
it was shown in [10] that (1.1) with Lagrangian $f=R \ln R$ describes
the same black hole space-time as the string inspired 2D dilaton gravity.

One of the motivations for recent investigations of 2D gravity (mainly
of the dilaton type) that it can be considered as a "toy"  model to study
the process of formation and subsequent evaporation of a black hole. It
has been argued by Hawking [11] that such a process is not governed by the
usual
laws of quantum mechanics: rather, pure states evolve into mixed states.
However, it is commonly believed that a successful quantization
of gravity and matter will provide us with a consistent solution of this
problem. Quantum corrections may completely change the gravitational
equations and the corresponding space-time geometry at the Planck
scales.
This problem is hard to analyze in four space-time dimensions. However,
in two dimensions one can attempt to attack this problem using the dilaton
gravity
theories as a toy model [3]. These toy models have an explicit semiclassical
treatment
of the back reaction of the Hawking radiation on the geometry of an
evaporating black hole by  including the one-loop Polyakov-Liouville
term in the action (the review can be found in [12]). Unfortunately, the
resulting equations are not exactly integrated
and one can not obtain a definite answer. Therefore, one can try to find
another theory of 2D gravity (among the alternatives) for which the relevant
semiclassical equations would be analytically solvable.

The main goal of our paper is the study of this problem for 2D  gravity
described by an action of the form (1.1) along the lines of ref.[3]. We show
that for
$f=R \ln R$ the semiclassical field equations are exactly integrated and
one can obtain a definite answer about the structure of
space-time when the backreaction of the Hawking radiation on the  black hole
geometry is taken into account.

This paper is organized as follows. In the next two sections we investigate
some
aspects common for theories described by the action (1.1). In Sec.2 we
demonstrate
the integrability of  classical field equations and find the exact
solution.
The one-loop renormalizability of the theory is analyzed in Sec.3. In the next
two
sections we mainly consider the case $f=R \ln R$. The coupling with conformal
(scalar)
matter is shown to be exactly solvable classically in Sec.4. The backreaction
is taken into account in Sec.5.

\section{Classical solution of the model}
\setcounter{equation}0

Under variation of the action (1.1) with respect to the metric $g_{\mu\nu}$ we
obtain
the following equations of motion:
\begin{equation}
\nabla_{\mu} \nabla_{\nu} [f']={1 \over 2} g_{\mu\nu} \{f(R)-Rf'(R) +2 \Box
(f') \}
\end{equation}
where $f' \equiv \partial_{R} f(R)$ and $\Box = \nabla^\mu \nabla_\mu$.

At first sight, (2.1) is a system of differential equations of very high order
with respect to derivatives. For example, if $f=R^2$, then (2.1) are equations
of fourth order of metric $g_{\mu\nu}$ derivatives. However, we will see that
it is not really so and the system (2.1) is rather easily solved.

Let us analyze at first  possible solutions of (2.1) with the constant
curvature
$R=R_{1}=const$. In this case we obtain that $f'(R)=const=f'|_{R=R_{1}}$
everywhere
in $M^2$. Then, from (2.1) we get that such a solution exists if the function
\begin{equation}
V(R)=f(R)-Rf'(R)
\end{equation}
is zero at the point $R=R_{1}$: $V(R_{1})=0$.  If $V(R)$ becomes zero at $P$
different points $R_{i}, i=1,2, ...,P$, then for given $f(R)$ there are $P$
different solutions of (2.1) with constant curvature. An additional
condition is that the function $f'(R)$ must be finite at $R=R_{i}$.

Assuming $R$ to be a non-constant  function on $M^2$, we consider a new
variable
$\phi=f'(R)$ provided that this equation is solved (at least locally) with
respect to $R$: $R=R(\phi)$. Denote $V(\phi) \equiv V(R(\phi))$.
Then (2.1) is rewritten as an equation on the new field $\phi$:
\begin{equation}
\nabla_{\mu}\nabla_{\nu} \phi = {1 \over 2} g_{\mu\nu} \{ V(\phi) +2 \Box \phi
\}
\end{equation}

We obtain from (2.3) that $\xi_{\mu}= \epsilon_{\mu}^{\ \nu} \partial_{\nu}
\phi$
is the Killing vector [13]. Consequently, the field $\phi$ can be chosen as one
of
the coordinates on $M^2$. Then, metric reads
\begin{equation}
ds^2=g(\phi)dt^2-{1 \over g(\phi)} d\phi^2
\end{equation}
{}From (2.3) we get that
\begin{equation}
\Box \phi =-V(\phi)
\end{equation}
For the metric (2.4) we have $\Box \phi =-g'(\phi)$ and eq.(2.5) reads
\begin{equation}
\partial_{\phi} g(\phi)=V(\phi)
\end{equation}
The solution  takes the form
\begin{equation}
g(\phi)= \lambda + \int\limits_{}^{\phi}V(\psi) d\psi
\end{equation}

One can see that our model (1.1), (2.1) seems to be equivalent  to some kind of
2D dilaton gravity with the dilaton field $\phi$ and potential $V(\phi)$.

Thus, surprisingly, our initial higher-derivative equations reduced to
the first order equation (2.6) independently of the concrete form of the
function
$f(R)$. As a result, the solution (2.7) is determined only by one arbitrary
integrating constant $\lambda$.
The Killing vector $\xi_\mu=\epsilon_\mu^{\ \nu} \partial_\nu \phi$
has bifurcation at a point where $\xi^2=-(\nabla \phi)^2$ equals zero.
One can see from (2.4) that $g(\phi)=0$ at this point and we have a horizon.

Since the scalar curvature for the metric (2.4) is equal to $R=-g''(\phi)$, one
can easily
 check that curvature for the solution (2.7) really coincides (if $f''(R)
\neq 0$) with $R=R(\phi)$ obtained by solving the equation $\phi=f'(R)$.

In the vicinity of points where $\phi'_{R}=f''(R)=0$ the equation $\phi=f'(R)$
cannot
be solved in a unique way. It is the only place where above solution  is
in-correct. Near
a point like that the function $\phi(R)$ is as shown in Fig.1 : there are two
values $R_{a},
R_{b}$ which correspond to the same value $\phi$. So there are two branches of
solution of equation $\phi=f'(R)$.  Let us consider this in more detail.
Let $\phi'(R)=0$ for some finite $R=R_{0}$. Note that only zero of odd order
is interesting for us.  In the vicinity of $R_{0}$ the function $\phi'(R)$ can
be represented as follows
\begin{equation}
\phi'(R)=a(R-R_{0})^{2k-1}, \ k=1,2,... \ , \ \ and \ \ a>0
\end{equation}
and hence we obtain
\begin{equation}
\phi(R)= {a \over 2k} (R-R_{0})^{2k} +b
\end{equation}
and
\begin{equation}
f(R)= {a \over 2k(2k+1)} (R-R_{0})^{2k+1} +bR+c
\end{equation}
There are two branches of the solution of eq.(2.9) with respect to $R$:
\begin{equation}
R=R_{0} \pm [(\phi -b) {2k \over a}]^{1 \over 2k}
\end{equation}
In the vicinity of the point $x \in M^2$ where $R(x)=R_{0}$ there are two
regions:
where $R>R_{0}$ and where $R<R_{0}$. Our solution (2.4), (2.7) is valid
in any of these regions taken separately. Consider the region where $R>R_{0}$.
Suppose for simplicity that $b=0, \ a>0$; then $\phi >0$. Then, we get for the
potential
$V$:
\begin{equation}
V(\phi)= -{2k \over 2k+1} ({2k \over a})^{1\over 2k} \phi^{1+ 1/2k} -R_{0} \phi
+c
\end{equation}

The corresponding metric (2.7) for $\phi >0$ reads
\begin{equation}
g(\phi)= \lambda -{R_{0} \over 2} \phi^2 +c \phi  - {(2k)^2 \over (2k+1)(4k+1)}
({2k \over a})^{1/2k} \phi^{2+1/2k}
\end{equation}
One can see that the metrical function $g(\phi)$ (2.13) has regular
in $\phi=0$
first and second derivatives

$$
g'(0)=0, \ g''(0)=-R_{0}
$$

However, the non-analyticity of (2.13) in $\phi=0$ manifests itself in that
all the
following derivatives are singular at this point:

$$
g^{(p)}(0)= \pm \infty , \ p>2
$$

This singularity means, in particular, that invariant $(\nabla R)^2$ is
singular at the point $x$ where $R(x)=R_{0}$. It should be noted that
singularities of
this type were earlier observed in [14,15] for rather different  theories.

In the region where $R<R_{0}$ we get
\begin{eqnarray}
&&R=R_{0}-( {2k \phi \over a})^{1/2k}  \nonumber \\
&&V(\phi)={2k \over 2k+1} ({2k \over a})^{1/2k} \phi^{1+1/2k} -R_{0} \phi +c,
\ \phi>0
\end{eqnarray}

Thus, the total space-time in the vicinity of the  point $R=R_{0}$ ($\phi=0$)
is represented
by gluing of two sheets (the coordinate $\phi>0$ can be used to
parameterize the points of both sheets in the neighborhood of $\phi=0$). The
total space-time
is shown in Fig.2.

Really, the scalar curvature $R$ itself can be used as one of the coordinates.
It covers, in particular, the whole vicinity of the point $R_{0}$. Then,  in
the coordinates
$(t,R)$ the metric reads:
\begin{equation}
ds^2=g(R)dt^2 - {[f''(R)]^2 \over g(R)} dR^2
\end{equation}
For $R \sim R_{0}$ we can put $g(r) \sim 1, \ f''(R) \sim a(R-R_{0})^{2k}$ and
hence the metric takes the form
\begin{equation}
ds^2=dt^2-a^2(R-R_{0})^{4k} dR^2
\end{equation}

Let us now consider some examples.

{\it Example 1.}

\begin{equation}
f(R)=R \ln R
\end{equation}
In this case $\phi= \ln R +1, \ R= e^{\phi -1}$. Hence $V(\phi)=-R=-e^{\phi
-1}$.
Since the potential $V(R)=-R$ is zero in $R=0$, it seems that one of the
solutions is flat
space-time. However, the function $f'(R)=\ln R +1$ is not defined for $R=0$. So
if we
come back to eq.(2.1), we observe that flat space-time is not really a solution
of the field equations.

If $R$ is a non-constant function on $M^2$, the solution is
given by the metric (2.4) with
\begin{equation}
g(\phi)= \lambda-R(\phi)=\lambda-e^{\phi -1}
\end{equation}
This solution coincides with that  obtained in 2d dilaton gravity and
describes asymptotically flat black hole space-time.
The essential difference of the solution (2.18) from that we have in
 dilaton gravity is that it doesn't describe flat space-time for any
integrating constant $\lambda$.

The Lagrangian (2.17) seems to be ill-defined at $R=0$. However,
we see that curvature $R(\phi)$ is everywhere positive and the point
$R=0$   really lies  at the spatial infinity.

{\it Example 2.}

\begin{equation}
f(R)=aR^2 +bR+c
\end{equation}
In this case $\phi=2aR+b, \ R={1 \over 2a} (\phi -b)$,i.e. $R(\phi)$ is linear
function. Then, we get $V(R)=-aR^2+c=- {1 \over 4a} (\phi -b)^2 +c$. If $c/a
>0$, then
$V(R)$ is zero at the points $R=\pm \sqrt{c/a}$. Thus, there are two solutions
with
the constant curvature: $R=\pm \sqrt{c/a}$. If $R$ is non-constant on $M^2$,
then
the solution is given by (2.4) with $g(\phi)$ in following form:
\begin{equation}
g(\phi)= \lambda+c \phi -{1 \over 12a} (\phi-b)^3
\end{equation}
This function has extremums at the points $\phi_{1,2}=b \pm 2a\sqrt{c/a}$
corresponding
to the curvature $R=\pm \sqrt{c/a}$.
Depending on the constant $\lambda$ (if $a,b,c$ are fixed), $g(\phi)$ can
have one, two or three zeros. It is worth noting that the space-time described
by
the metric (2.20) is not asymptotically flat. $R=0$ is reached at the point
$\phi=b$ which
stays on finite distance from any point $\phi \neq \pm \infty$. Thus,
the points $\phi =\pm \infty$ lies at asymptotical infinity and the curvature
is
singular at this point. In this sense, the solution (2.20) is similar to that
obtained
in the 2D theory of gravity with torsion described by the action quadratic in
curvature
and torsion [5].

It should be noted that a behavior like that is rather typical of
polynomial gravity (1.1). Indeed, let the function $f(R)$ near $R=0$ look like
as
$f(R)=aR^{\alpha} +c$. Then, $\phi=\alpha a R^{\alpha-1}$, $R(\phi)=({\phi
\over \alpha a})^{
1/{\alpha -1}}$. One can see that for $\alpha>1$ the point $R=0$ corresponds to
$\phi=0$ and,
consequently, lies at a finite distance from any point $\phi \neq 0$. It means
that space-time
is not asymptotically flat.  The last is reached only if $\alpha<1$: then $R
\rightarrow 0$
means $\phi \rightarrow +\infty$. However, in this case the function $f(R)$ is
not analytical
in $R=0$. It is the case in ${\it Example 1.}$, where the solution is
asymptotically flat.
More generally, the solution is asymptotically flat if the function $f(R)$
satisfies
the condition: $f'(R) \rightarrow \pm \infty$ if $R \rightarrow 0$.
It is easy to see, however, that flat space-time is not a solution of field
equations
in this case, as we have seen in ${\it Example 1.}$

\section{One-loop renormalization}
\setcounter{equation}0

The complete quantization of the model (1.1) is a rather difficult problem.
In this section, we just calculate  one-loop counter-terms and
check the renormalizability of the model in one loop and  not
considering these problems as the unitarity.
We assume in this section that $f''(R) \neq 0$.

We use the background method. The metric $g_{\mu\nu}$ is written in the form:
$g_{\mu\nu}=\bar{g}_{\mu\nu}+h_{\mu\nu}$, where $\bar{g}_{\mu\nu}$ is a
classical background metric, $h_{\mu\nu}$ is a small quantum field.
In the conformal gauge we have $h_{\mu\nu}=\sigma /2 \bar{g}_{\mu\nu}$ and
the theory reduces to quantization of only a conformal mode $\sigma$.
Expanding the action (1.1) in powers of $\sigma$ we obtain the quadratic in
$\sigma$
expression
\begin{eqnarray}
&&S[g_{\mu\nu}]=S_{cl}[\bar{g}_{\mu\nu}]+ S_{q}[\sigma] \nonumber \\
&&S_{q}[\sigma]=\int\limits_{M^2}^{}[f''(R) ( \Box \sigma)^2
-2Rf''(\sigma \Box \sigma) +(R^2 f'' +f-Rf')\sigma^2] \sqrt{\bar{g}}d^2 z
\end{eqnarray}
where the curvature $R=R[\bar{g}]$ and the Laplacian $\Box = {1 \over
\sqrt{\bar{g}}}
\partial_{\mu}[\sqrt{\bar{g}}\bar{g}^{\mu\nu} \partial_{\nu}]$ are determined
with respect to the background metric $\bar{g}_{\mu\nu}$.
We see from (3.1) that $(1/f'')$ is effectively the loop expansion parameter
for gravity.

The action $S_{q}[\sigma]$ can be written in the form
\begin{equation}
S_{q}[\sigma]= \int\limits_{M^2}^{}
\sqrt{\bar{g}}d^2 z
\sigma \hat{D} \sigma
\end{equation}
where $\hat{D}$ is the fourth-order differential operator
\begin{equation}
\hat{D}=(\Box +X)f''(\Box +Y)
\end{equation}
and the functions $X$, and $Y$ satisfy  the following equations:
\begin{equation}
X+Y=-2R, \ XY=R^2 f''+f-Rf'
\end{equation}
Calculating the functional integral over the conformal factor $\sigma$ we can
compute
the infinite part of the one-loop effective action
\begin{equation}
I_{\infty}={1 \over 2} (\ln det \hat{D})_{\infty} -(\ln det
\Delta_{gh})_{\infty}
\end{equation}
where $\Delta_{gh}$ is the standard ghost operator corresponding to the
conformal
gauge.

By definition, for an elliptic $2r$ order differential operator
$\Delta$ defined on a two-dimensional Riemannian manifold $M^2$ we get:
\begin{equation}
\ln det \Delta =-\int\limits_{\epsilon}^{+\infty} {dt \over t} Tr e^{-t
\Delta}, \
\epsilon \rightarrow +0
\end{equation}
The infinite part is given by ($L \rightarrow \infty$):
\begin{equation}
(\ln det \Delta )_{\infty}=- (B_{0} L^2 +2B_{1}L +{r \over 2} B_{2} \ln {L^2
\over \mu^2})
\end{equation}
where
\begin{eqnarray}
&&(Tr e^{-t \Delta })_{t \rightarrow 0} = \sum_{k=0}^{2} B_{k} t^{(k-2)/2r} +
O(\sqrt{t}),
\nonumber \\
&&B_{k}=\int\limits_{M^2}^{} b_{k}(\Delta) \sqrt{g}d^2 z+
\int\limits_{\partial M^2}^{}c_{k}(\Delta) \sqrt{\gamma}d\tau,
\end{eqnarray}
$b_{k}(\Delta)$ are the Seeley coefficients for the operator $\Delta$
($b_{2p+1}=0$).
For simplicity we will assume that $M^2$ is a manifold without a boundary
($\partial M^2=0$) and will neglect all boundary effects.

Note that $L^2$ and $L$ dependent terms are automatically absent in the
dimensional
or $\zeta$-function regularization. So only the last term in (3.7) is
of  interest for us.

Now consider the Seeley coefficient for the elliptic fourth-order operator
$\hat{D}$ (3.3) (see [16]). Suppose that for some $\Delta_{4}$
\begin{equation}
\Delta_{4}=\Delta_{2} \Delta'_{2} , \ det \Delta_{4}=det \Delta_{2} det
\Delta'_{2}
\end{equation}
Then, we get the corresponding expression for the Seeley coefficients [16]:
\begin{equation}
2B_{2}(\Delta_{4})=B_{2}(\Delta_{2}) + B_{2}(\Delta'_{2}),
\end{equation}
Since the operator $\hat{D}$ (3.3) has the structure (3.9)
we obtain that
\begin{equation}
2B_{2} (\hat{D})=
B_{2} (\Box +X) +
B_{2} (f''(\Box +Y))
\end{equation}
For the operator
\begin{equation}
\Delta_{2}= \Box +X
\end{equation}
the following result is well-known:
\begin{equation}
b_{0}(\Delta_{2})={1 \over 4\pi} \ ; \ b_{2}(\Delta_{2})={1 \over 4\pi} (1/6
R+X)
\end{equation}

To calculate the Seeley coefficients for the operator
\begin{equation}
\Delta'_{2}=f''(\Box +Y),
\end{equation}
it is useful to observe that this operator can be transformed to
 (3.12) by introducing a new metric $\tilde{g}_{\mu\nu}=(f'')^{-1} g_{\mu\nu}$:
\begin{equation}
\Delta'_{2}=\Box_{\tilde{g}}+ f'' Y,
\end{equation}
where $\Box_{\tilde{g}}= { 1 \over \sqrt{\tilde{g}}}
\partial_{\mu}[\sqrt{\tilde{g}}\tilde{g}^{\mu\nu} \partial_{\nu}]$.
Now using the result (3.13) we obtain
\begin{equation}
B_{2}(\Delta'_{2})={1 \over 4\pi} \int\limits_{M^2}^{}( 1/6 \tilde{R} +f'' Y)
\sqrt{\tilde{g}}d^2 z
\end{equation}
or in terms of the old metric $g_{\mu\nu}$ eq.(3.16) takes the form
\begin{equation}
B_{2}(\Delta'_{2})={1 \over 4\pi} \int\limits_{M^2}^{}( 1/6 R + Y)
\sqrt{g}d^2 z
\end{equation}
Thus, we obtain for $B_{2}(\hat{D})$:
\begin{equation}
B_{2}(\hat{D})={1 \over 8\pi} \int\limits_{M^2}^{}( 1/3 R + X+Y)
\sqrt{g}d^2 z
\end{equation}
Using (3.4) we finally get
\begin{equation}
B_{2}(\hat{D})=-{1 \over 4\pi} \int\limits_{M^2}^{} 5/6 R
\sqrt{g}d^2 z
\end{equation}

The Seeley coefficient for the ghost operator $\Delta_{gh}$ is well-known [17]
\begin{equation}
 b_{2}(\Delta_{gh})={ 1\over 4\pi}( 2/3 R)
\end{equation}
Taking into account (3.7), (3.19-20) we obtain from (3.5)
that the corresponding one-loop counter-term
\begin{equation}
I_{ct}=a \int\limits_{M^2}^{} R \sqrt{g}d^2 z
\end{equation}
is surprisingly non-dependent on the concrete function
$f(R)$. So the model (1.1) seems to be one-loop renormalizable.
Of course, if one takes the next loops this result could be changed
and possibly not for any $f(R)$ the theory is renormalizable.
However, we do not consider higher loops here.

\bigskip

\section{Coupling with conformal matter}
\setcounter{equation}0

Let us consider interaction of higher derivative gravity (1.1)
with 2d conformal matter described by the action
\begin{equation}
S_{mat}=\int\limits_{M^2}^{} {1 \over 2} (\nabla \psi)^2 \sqrt{g} d^2 z
\end{equation}

Then, we get the complete system of equations of motion:
\begin{equation}
T_{\mu\nu} \equiv \nabla_{\mu} \nabla_{\nu} \phi -{1 \over 2}
g_{\mu\nu} [V(\phi) +2 \Box \phi] +{1 \over 2} (\partial_{\mu}
\psi  \partial_{\nu} \psi -{1 \over 2} g_{\mu\nu} \partial_{\alpha} \psi
\partial^{\alpha} \psi)=0
\end{equation}
where $\phi=f'(R)$. The equation of motion for matter reads
\begin{equation}
\Box \psi=0
\end{equation}
We will use the conformal gauge in which the components of metric:
$g_{++}=g_{--}=0; \ g_{+-}= {1 \over 2} e^{\sigma}$. Then, eq.(4.2) takes the
form
\begin{equation}
T_{\pm \pm}=\partial_{\pm} \partial_{\pm} \phi -\partial_{\pm}
\sigma \partial_{\pm}\phi +{1 \over 2} \partial_{\pm} \psi \partial_{\pm} \psi
=0
\end{equation}
\begin{equation}
T_{+-}=0 \ <=> \ 4\partial_{+} \partial_{-} \phi =-V(\phi) e^{\sigma}
\end{equation}
Moreover, we have the self-consistency condition :
\begin{equation}
4\partial_{+} \partial_{-} \sigma =R(\phi) e^{\sigma}
\end{equation}
Equation (4.3) takes the form

$$
\partial_{+}\partial_{-} \psi =0
$$
and the solution reads
\begin{equation}
\psi=\psi_{+} (x^{+}) + \psi_{-} (x^{-})
\end{equation}

For the function $f(R)$ of general form  eqs.(4.4-6) are
extremely non-linear  differential equations which are not
exactly solved in general. However, in some particular cases, for concrete
$f(R)$, this problem can be essentially simplified.

Let us consider the case when the following equation is valid:
\begin{equation}
\partial_{+}\partial_{-} (\sigma -\phi)=0
\end{equation}
It is the case when $f(R)$ satisfies the equation:
\begin{equation}
R+V(R)=R+f-Rf'=0,
\end{equation}
i.e., when $f(R)=R \ln R$. This case was considered in ${\it Example 1.}$

Then, we get
\begin{equation}
\sigma-\phi= w_{+} (x^{+}) +w_{-} (x_{-})
\end{equation}
On the other hand, one can see that $(\sigma + \phi)$ satisfies the Liouville
equation
\begin{equation}
\partial_{+} \partial_{-} (\sigma + \phi)={1 \over 2e} e^{\sigma + \phi}
\end{equation}
which has following general solution:
\begin{equation}
\sigma + \phi = \ln {A'(x^+) B'(x^-) \over [1-{1 \over 4e} AB]^2}
\equiv \beta(x^+,x^-),
\end{equation}
where $A$ and $B$ are still unknown functions of $x^+$ and $x^-$
respectively.

Thus, we obtain:
\begin{equation}
\sigma={1 \over 2} \beta + {1 \over 2} w, \ \phi ={1 \over 2} \beta - {1 \over
2}
w
\end{equation}
{}From this we see that

$$
\partial^2_{+} \phi - \partial_{+} \sigma \partial_{+} \phi=
{1 \over 2} (\partial^2_{+} \beta -1/2 (\partial_{+}\beta)^2)-
{1 \over 2} (\partial^2_{+}w  -1/2 (\partial_{+} w)^2)
$$
On the other hand, one can see the following identity:
$$
(\partial^2_{+} \beta -1/2 (\partial_{+}\beta)^2)=
\{A; x^+ \}
$$
where we introduced the Schwarzian derivative
\begin{equation}
\{  {\cal F}; x \}= { \partial^{3}_{x} { \cal F} \over \partial_{x} {\cal F}} -
{3 \over 2} ( {\partial^{2}_{x} {\cal F} \over \partial_{x} {\cal F}})^2
\end{equation}
Then, we get for the $(++)$-component of equation (4.4):
\begin{equation}
\{A;x^+ \} -(\partial^2_{+} w_{+} -1/2 (\partial_{+} w_{+})^2)+
2T_{++}^{\psi}=0
\end{equation}
where $T^{\psi}_{++}=1/2 (\partial_{+} \psi_{+})^2$. Similarly,
we obtain for the $(--)$-component of eq.(4.4):
\begin{equation}
\{B;x^- \} -(\partial^2_{-} w_{-} -1/2 (\partial_{-} w_{-})^2)+
2T_{--}^{\psi}=0
\end{equation}
Using the known property of the Schwarzian derivative (see for example [18]),
one can see that eqs.(4.15)-(4.16)
are invariant under $SL(2,R) \oplus SL(2,R)$ group transformations:
\begin{eqnarray}
&&A \rightarrow {a A+b \over c A+d}, \ ad-bc=1 \nonumber \\
&&B \rightarrow {m B+n \over k B+p}, \ mp-kn=1
\end{eqnarray}
Under the coordinate transformations $x^{\pm} \rightarrow y^{\pm}(x^{\pm})$ we
have:
\begin{equation}
\beta(x^{+}, x^{-}) \rightarrow \beta(y^{+}, y^{-}) -\ln ({\partial y^{+} \over
\partial x^{+}} { \partial y^{-} \over \partial x^{-}})
\end{equation}
On the other hand, $w_{\pm}$ transforms as follows:
\begin{equation}
w_{\pm}(x^{\pm}) \rightarrow w_{\pm}(y^{\pm}) - \ln ({\partial y^{\pm} \over
\partial x^{\pm}})
\end{equation}
We can use this symmetry to put  $w_{\pm}=0$. Then, one obtains equations
on functions $A$ and $B$:
\begin{eqnarray}
&&\{ A; x^+ \} = -(\partial_{+} \psi_{+})^2 \nonumber \\
&&\{ B; x^- \} = -(\partial_{-} \psi_{-})^2
\end{eqnarray}
When the matter is absent $(T_{\pm \pm}=0)$, the solution of equations
\begin{equation}
\{A; x^+ \}=0, \ \{B; x^- \}=0
\end{equation}
is one of the following types. If $A''=0$, then
\begin{equation}
A=ax^{+} +b
\end{equation}
if $A'' \neq 0$, then
\begin{equation}
A=f- {a \over x^{+}+b}
\end{equation}
Correspondingly, we get for $B(x^{-})$:
\begin{equation}
B=m x^{-} +n
\end{equation}
or
\begin{equation}
B=d- {m \over x^{-} +n}
\end{equation}
The metric takes the form:
\begin{equation}
ds^2={[A'B']^{1/2} \over (1-{AB \over 4e})} dx^{+} dx^{-}
\end{equation}
Shifting $x^{+}, x^{-}$ on constants we get $b=n=0$ in (4.22-25).
Though $A,B$ depend on the set of constants, the metric (4.26) depends only on
one arbitrary
constant.
Let, for example, $A''=B''=0$, then
\begin{equation}
ds^2= {c
dx^{+}dx^{-}
\over (1- {c^2 \over 4e} x^{+}x^{-})}
\end{equation}
where $c= \sqrt{am}$.
In other cases, if $A''$ and $B'' $ are not zero, the metric takes the form:
\begin{equation}
 ds^2= {c dx^{+}dx^{-} \over c^2 x^{+}x^{-} - {1 \over 4e}}
\end{equation}
where $c=(1-{fd \over 4e}) (am)^{-1/2}$.
The scalar curvature is given by the formula:
\begin{equation}
R={1 \over e} {(A'B')^{1/2} \over (1-{AB \over 4e})}
\end{equation}
It has singularity if $AB=4e$, which one can also see from eqs.(4.27), (4.28).
The points of horizon satisfy: $AB=0$.
The space-time (4.27-28) is of the same type as the black hole solution in
the 2D dilaton gravity [1,3]. However,
there is no such integrating constant for which the metric (4.27-28) is  flat.
The flat space-time is not a solution of field equations that has
already been noted above. This is an essential difference between the string
inspired 2D
dilaton gravity [1] and higher derivative gravity (1.1) with $f=R \ln R$.
Therefore, eqs.(4.4)-(4.6) do not describe the black hole formation
from regular (flat) space-time due to the infalling matter as we had in
the dilaton gravity [3]. The "bare" black hole is necessary. The infalling
matter only deforms this initially singular space-time.

\bigskip

As an example let us now consider the falling of $\delta$-like impulse of
matter on
the black hole.  The matter energy-momentum tensor takes the form:
$T^{\psi}_{++}= \lambda \delta (x^{+}-x^{+}_{0}), \ T^{\psi}_{--}=0$
$(\lambda>0)$.
It describes the $\delta$-like impulse of matter propagating along
the $x^{-}$-direction. Suppose that the space-time for $x^{+}<x^{+}_{0}$
is a solution of the field equations without matter such that $A=ax^{+}, \ B=m
x^{-}$.
For $x^{+}>x^{+}_{0}$ the function $B(x^-)$ is the same  while $A(x^{+})$ is
found
from the equation:
\begin{equation}
\{A; x^{+} \}=- \lambda \delta(x^{+}-x^{+}_{0})
\end{equation}
For $x^{+}>x^{+}_{0}$ the function $A(x^{+})$ is a solution of eq.(4.30) with
the zero
right-hand side
\begin{equation}
A={\alpha x^{+}+\beta \over \kappa x^{+} +\gamma}, \ \alpha\gamma-\kappa\beta=1
\end{equation}
where the constants $\alpha,\beta,\kappa,\gamma$ are found from the continuity
condition
of functions $A(x^{+})$,
$A'(x^{+})$ and the gap
condition for $A''(x^{+})$ at the point $x^{+}=x^{+}_{0}$. The last condition
is
easily obtained integrating (4.30) in the interval $(x^{+}_{0}- \epsilon,
x^{+}_{0}+\epsilon)$ and then taking the limit $\epsilon \rightarrow 0$. As a
result
one obtains:
\begin{equation}
A''(x^{+}_{0}+0)-A''(x^{+}_{0}-0) =-\lambda A'(x^{+}_{0})
\end{equation}
{}From continuity of $A(x^{+})$  and $A'(x^{+})$ one gets:
\begin{equation}
ax^{+}_{0}={\alpha x^{+}_{0} +\beta \over \kappa x^{+}_{0} +\gamma}
\end{equation}
\begin{equation}
a=(\kappa x^{+}_{0} +\gamma)^{-2}
\end{equation}
and from the gap condition (4.32) we obtain
\begin{equation}
{2\kappa \over (\kappa x^{+}_{0} +\gamma)}=\lambda
\end{equation}
These equations and $\alpha\gamma-\kappa\beta=1$ are enough to find the form
of $A(x^{+})$ for $x^{+}>x^{+}_{0}$:
\begin{equation}
A(x^{+})=a {x^{+} + {\lambda x^{+}_{0} \over 2} (x^{+}-x^{+}_{0}) \over
1+{\lambda \over 2}(x^{+}-x^{+}_{0})}
\end{equation}
The metric for $x^{+}<x^{+}_{0}$ takes the form
\begin{equation}
ds^2= { \sqrt{am}dx^{+}dx^{-} \over (1- {am \over 4e} x^{+}x^{-})}
\end{equation}
and the corresponding curvature is the following
\begin{equation}
R=e^{-1} {\sqrt{am} \over (1-{am \over 4e} x^{+}x^{-})}
\end{equation}
This metric describes the black hole with horizon in $x^{+}x^{-}=0$ and
singularity
at $x^{+}x^{-}={ 4e \over am}$.

We will assume that $x^{+}_{0}>0$, i.e. impulse  falls from asymptotically
flat region which lies right of horizon $(x^{+}=0)$. Then, for
$x^{+}>x^{+}_{0}$
we obtain for the metric
\begin{equation}
ds^2= {\sqrt{am} \over [1+{\lambda \over 2}(x^{+}-x^{+}_{0})]}
{dx^{+} dx^{-} \over [1-{am \over 4e} ({x^{+}+ {\lambda x^{+}_{0} \over 2}
(x^{+}-x^{+}_{0}) \over 1+ {\lambda \over 2}(x^{+}-x^{+}_{0})})x^{-}]}
\end{equation}
and the curvature
\begin{equation}
R=e^{-1} \sqrt{am}[1+{\lambda \over 2} (x^{+}-x^{+}_{0}) -{am \over 4e}
(x^{+}+ {\lambda x^{+}_{0} \over 2}(x^{+}-x^{+}_{0})) x^{-}]^{-1}
\end{equation}

One can see from (4.40) that for $x^{+}>x^{+}_{0}$ singularity lies
on the curve:
\begin{equation}
x^{-}={4e \over am} {(1+{\lambda \over 2}(x^{+}-x^{+}_{0})) \over
(x^{+}+{\lambda x^{+}_{0} \over 2}(x^{+}- x^{+}_{0}))}
\end{equation}
The derivative of the function
(4.41):

$$
\partial_+ x^-=-{4e \over am} (x^++{\lambda x^+_0 \over 2} (x^+-x^+_0))^{-2}
$$
is negative and we have that for $x^+>x^+_0$ the function (4.41) is the
monotonically
decreasing one
 smoothly glued with  $x^{-}={4e \over am} {1 \over x^{+}}$ at
$x^{+}=x^{+}_{0}$. Moreover, in the limit $x^{+} \rightarrow \infty$ it
limits to $x^{-} \rightarrow x^-_{\infty}={4e \over am} (x^{+}_{0}+2/
\lambda)^{-1}$.
The total space-time for all $x^{+}$ is shown in Fig.3.
In the asymptotically flat region ($x^+>0$) it is similar to that we have  for
the
2D dilaton gravity case [3].

\bigskip

It should be noted that the function $f(R)=R \ln R$ is not a unique one for
which
equations (4.4-6) are exactly integrated. Indeed, we obtain from
(4.5-6):
\begin{eqnarray}
&&4\partial_{+}\partial_{-} (\sigma-\phi)=(R+V)e^{\sigma} \nonumber \\
&&4\partial_{+}\partial_{-} (\sigma+\phi)=(R-V)e^{\sigma}
\end{eqnarray}
These equations are reduced to the system of the Liouville equations if
\begin{equation}
R+V=ae^{-\phi}, \ R-V=b e^{\phi}
\end{equation}
where $a,b$ are constants. These conditions are equivalent to the system of
differential equations on the function $f(R)$
\begin{eqnarray}
&&R+f-Rf'=ae^{-f'} \nonumber \\
&&R-f+Rf'=be^{f'}
\end{eqnarray}
One obtains immediately from this
\begin{equation}
R=1/2[ae^{-f'}+be^{f'}]
\end{equation}
Solving (4.45) with respect to $f'$ one obtains:
\begin{equation}
f'=\ln {R \pm   \sqrt{R^2-ab} \over b}
\end{equation}
Integrating this we finally get:
\begin{equation}
f(R)=R \ln {R \pm \sqrt{R^2-ab} \over b} \mp \sqrt{R^2-ab}
\end{equation}
where  both signs ($\pm$) are available if $(ab)>0$. We do not consider here
this kind
of theory. Note only that for $a, b \neq 0$ it describes the
asymptotically singular rather than asymptotically flat space-time.

\bigskip

\section{Solution with backreaction}
\setcounter{equation}0
As we have discribed in the Introduction
 quantum corrections are usually assumed to remove the black hole
singularity.
One can try to analyze this problem semiclassically considering quantum gravity
coupled
to a large number $N$ of free scalar fields. In the limit $\hbar \rightarrow 0$
with $N \hbar$ fixed, it is a system in which the leading order of
a perturbative expansion is a quantum theory of matter in classical geometry.
Integrating out the matter we have an effective action describing the
backreaction
of matter and Hawking radiation on the geometry, which we  hope to treat
classically [19]. In ref.[3] it was proposed to use this approach to study the
problem
in two dimensions for dilaton gravity. However, the resulting quantum-corrected
field equations
are not exactly solved [3,12,20] though some reasons observed in favor
of that singularity are still present in this semiclassical theory.
We apply here the approach of [3] for theory of gravity described by the action
(1.1).

In two dimensions, integrating out the conformal scalar fields one
gets the Polyakov-Liouville action:
\begin{equation}
S_{PL}={N \over 96 } \int\limits_{}^{}d^2 x_{1}
\sqrt{-g}\int\limits_{}^{} d^2 x_{2} \sqrt{-g} R(x_{1}) \Box^{-1}
(x_{1},x_{2}) R(x_{2})
\end{equation}
here $\Box^{-1}$ denotes the Green function for
the Laplacian. It should be noted that $S_{PL}$ incorporates both the
Hawking radiation and the effects of its backreaction on the geometry.
We neglect here the contribution of the ghosts [21].
The full effective action
\begin{equation}
S_{eff}=S_{gr}+S_{mat}+S_{PL}
\end{equation}
gives rise to the following system of equations (the metric is taken to
be conformally flat $g_{-+}={1 \over 2} e^\sigma$):
\begin{equation}
\partial_{\pm} \partial_{\pm} \phi -\partial_{\pm}
\sigma \partial_{\pm}\phi +
2c[\partial^{2}_{\pm} \sigma-{1 \over 2} (\partial_{\pm} \sigma)^2
-t_{\pm}]+T^{\psi}_{\pm\pm}=0
\end{equation}

\begin{equation}
4\partial_{+}\partial_{-} \phi=-e^\sigma(V(\phi)+2cR)
\end{equation}
where $c={N \over 48}$ and $T^\psi_{\pm\pm}={1 \over 2} \partial_\pm \psi
\partial_\pm \psi$. Equation (5.4) is obtained as variation of
the action (5.2) with respect to $g_{+-}$. Since the scalar curvature is
a known function of $\phi$, we must add the condition of self-consistency:
\begin{equation}
4\partial_{+}\partial_{-}\sigma=R(\phi)e^\sigma
\end{equation}
The scalar matter equation

$$
\partial_+ \partial_- \psi=0
$$
gives
$$
\psi=\psi_+(x^+)+\psi_-(x^-)
$$

\bigskip

For general function $f(R)$ these equations seem to be not exactly
integrated. Therefore, we will consider in this section only the case

$$
S_{gr}=\int_{}^{}R \ln R
\sqrt{-g} d^2x
$$
and show that for this type of gravitational action
the system (5.3-5) is exactly solved.
In this case $R(\phi)=e^{\phi-1}, \ V(\phi)=-R$. Equations
(5.4), (5.5) take the form
\begin{equation}
4\partial_{+}\partial_{-} \phi={(1-2c) \over e} e^{\phi+\sigma}
\end{equation}
\begin{equation}
4\partial_{+}\partial_{-} \sigma={1 \over e} e^{\phi+\sigma}
\end{equation}

\bigskip

{\it Let $c \neq 1$}.

 Then, from (5.6-7) we obtain
\begin{eqnarray}
&&\partial_{+}\partial_{-}[(1-2c)\sigma-\phi]=0 \nonumber \\
&&\partial_{+}\partial_{-}[\phi+\sigma]={(1-c) \over 2e} e^{\phi+\sigma}
\end{eqnarray}
These equations are easily solved
\begin{eqnarray}
&&(1-2c)\sigma-\phi=w_{+}(x^+)+w_{-}(x^-) \equiv w \nonumber \\
&&[\phi+\sigma]=\ln {A'B' \over (1-{(1-c) \over 4e} AB)^2} \equiv \beta
\end{eqnarray}
where $A=A(x^+), \ B=B(x^-)$.
Finally, we get for the conformal factor $\sigma$ and field $\phi$:
\begin{equation}
\sigma={1 \over 2(1-c)}(w+\beta); \ \phi={(1-2c) \over 2(1-c)}\beta-
{1 \over 2(1-c)} w
\end{equation}

With respect to the coordinate changing $x^{\pm} \rightarrow y^{\pm}
(x^{\pm})$ we have
\begin{eqnarray}
&&\beta(x^+,x^-) \rightarrow \ln \beta (y^+,y^-) +(\partial_{+}y^+ \partial_{-}
y^-)
\nonumber \\
&&w^{\pm}(x^{\pm}) \rightarrow w^{\pm}(y^{\pm})-(1-2c) \ln
(\partial_{\pm}y^{\pm})
\end{eqnarray}
Hence, the fields $\sigma$ and $\phi$ transform as a usual conformal factor
and a scalar field, respectively:
\begin{equation}
\sigma (x^+,x^-) \rightarrow \sigma (y^-,y^+) - \ln (\partial_{-}y^-
\partial_{+} y^+) \ ,
\phi(x^+,x^-) \rightarrow \phi (y^+,y^-)
\end{equation}
One can easily  see that
\begin{eqnarray}
&&\partial^{2}_{\pm}  \phi -\partial_{\pm}
\sigma \partial_{\pm}\phi +
2c[\partial^{2}_{\pm}\sigma -{1 \over 2} (\partial_{\pm} \sigma)^2] =
\nonumber \\
&&{1 \over 2(1-c)} [\partial^{2}_{\pm} \beta -{1 \over 2}
 (\partial_{\pm} \beta)^2]  +{1 \over 2(1-c)} [\partial^{2}_{\pm} w
 (-1+2c)
+{1 \over 2} (\partial_{\pm} w)^2]
\end{eqnarray}
As before, we have in terms of the Schwarzian derivative
\begin{eqnarray}
&&\partial^2_{+}\beta-{1 \over 2}(\partial_{+} \beta)^2= \{A;x^+ \}
\nonumber \\
&&\partial^2_{-}\beta-{1 \over 2}(\partial_{-} \beta)^2= \{B;x^- \}
\end{eqnarray}

\bigskip

Let moreover $c \neq 1/2$, then we can use the symmetry (5.11) to put $w=0$.
Then,
equations (5.3) take the form
\begin{eqnarray}
&&\{ A; x^+\} =-2(1-c)T^{\psi}_{++}+4c(1-c)t_{+}(x^+) \nonumber \\
&&\{ B; x^- \} =-2(1-c)T^{\psi}_{--}+4c(1-c)t_{-}(x^-)
\end{eqnarray}
Eqs.(5.15) are ordinary differential equations  with respect  to $A, \ B$.

The metric and curvature, respectively, read:
\begin{equation}
ds^2=e^{\beta \over 2(1-c)} dx^+dx^-=
[{A'B' \over (1-{(1-c) \over
4e}AB)^2}]^{1 \over 2(1-c)}
dx^+ dx^-
\end{equation}
\begin{equation}
R={ 1 \over e} e^{(1-2c)\beta \over 2(1-c)}= { 1 \over e}
[{A'B' \over (1-{(1-c) \over
4e}AB)^2}]^{(1-2c) \over 2(1-c)}
\end{equation}

If the right-hand side of eq.(5.15) is zero (i.e., matter is absent), then the
solution of eq.(5.15) is already known (see (4.22-25)). Let, for example,
it take the form (4.22), (4.24) with $b=n=0$: $A=ax^+, \ B=mx^-$. Then,
the metric and scalar curvature take the form
\begin{equation}
ds^2=
[{ \sqrt{am} \over (1-{(1-c)am \over 4e} x^+ x^-)}]^{ 1 \over 1-c}
dx^+ dx^-
\end{equation}
\begin{equation}
R={1 \over e}
[{ \sqrt{am} \over (1-{(1-c)am \over 4e} x^+ x^-)}]^{ ({1-2c \over 1-c}
)}
\end{equation}
If $c=0$, we obtain the "classical" black hole space-time with space-like
singularity at $x^+x^-={ 4e \over am}$ (we assume that $am>0$).

For $c<1/2$ or $c>1$ we see from (5.19) that space-time still has singularity
at
$x^+x^-={ 4e \over am(1-c)}$. It is time-like for $c>1$ and space-like for
$c<1/2$.
As before, the points of horizon satisfy the condition $(\partial \phi)^2=0$,
which for (5.18), (5.19) means that $x^+x^-=0$.
 The  diagram of this space-time is shown
in Fig.4. The regions I and III are asymptotically flat.

\bigskip

Let $c>1$ and consider the falling in this space-time of the matter impulse
$T^{\psi}_{++}=\lambda \delta (x^+-x^+_0), \ \lambda >0$. We will neglect
contribution of
$t_{\pm}(x^{\pm})$ into (5.15). We assume  that impulse falls in the region
I which is asymptotically flat, i.e. $x^+_0<0$. For $x^+ <x^+_0$ the space-time
is described by the metric (5.18) and has curvature (5.19). For $x^+>x^+_0$
the solution of eq.(5.15) is found in the same way as before (see the previous
section). Moreover, the solution has the form similar to (4.36)
\begin{equation}
A(x^{+})=a {x^{+} + {\lambda (1-c)x^{+}_{0} \over 2} (x^{+}-x^{+}_{0}) \over
1+{\lambda (1-c)\over 2}(x^{+}-x^{+}_{0})}, \ x^+>x^+_0
\end{equation}
We obtain correspondingly for the metric
\begin{eqnarray}
&&ds^2=(am)^{-1 \over 2(c-1)}
 [1+ {\lambda \over 2}(1-c)(x^+-x^+_0) -  \nonumber \\
&&{am(1-c) \over 4e} x^-(x^+ +{\lambda \over 2} (1-c) x^+_0(x^+-x^+_0))]^{1
\over c-1}
dx^+dx^-
\end{eqnarray}
and scalar curvature
\begin{eqnarray}
&&R=1/e (am)^{2c-1 \over 2(c-1)}
 [1+ {\lambda \over 2}(1-c)(x^+-x^+_0) - \nonumber \\
&&{am(1-c) \over 4e} x^-(x^+ +{\lambda \over 2} (1-c)
x^+_0(x^+-x^+_0))]^{-{2c-1 \over c-1}}
\end{eqnarray}
For $x^+>x^+_0$ the singularity  lies on the curve
\begin{equation}
x^{-}={4e \over am(1-c)} {(1+{\lambda (1-c)\over 2}(x^{+}-x^{+}_{0})) \over
(x^{+}+{\lambda (1-c)x^{+}_{0} \over 2}(x^{+}- x^{+}_{0}))}
\end{equation}
which is smoothly glued with the curve $x^-={4e \over am(1-c)} 1/x^+$
at the point $x^+=x^+_0$. Calculating a derivative of the function (5.23), we
obtain that
\begin{equation}
\partial_+ x^-=-{4e \over am(1-c)}(x^++{\lambda (1-c) \over 2} x^+_0
(x^+-x^+_0))^{-2}
\end{equation}
i.e., the function (5.23) is monotonically increasing  (remember that
we consider the case $c>1$). The function (5.23) takes an infinite value at
$x^+_1=(1+{2 \over \lambda (1-c) x^+_0})x^+_0$, $x^+_0 <x^+_1 <0$. It means
that singularity in the region II is slightly shifted for $x^+>x^+_0$,
as is shown in Fig.5. On the other hand, the singularity in the region
IV asymptotically tends to $x^-_\infty ={2e\lambda \over am} (1+{\lambda(1-c)
\over 2}
x^+_0)^{-1}$. The resulting space-time is shown in Fig.5.
We see that for large $N$ , $c>1$,
the singularity doesn't disappear but simply becomes  time-like, which is
similar to that we
have for the 2D dilaton gravity [20].

\bigskip

Another case happens if $c$ lies in the interval $1/2 <c<1$. One can see that
power
in the expression for the curvature (5.19) becomes negative. Hence, the metric
(5.18) describes space-time which is regular for any finite $x^+$ and $x^-$.
In particular, it is the case for the points on the line $x^+x^-= {4e \over
am(1-c)}$
(or $AB={4e \over (1-c)}$). The curvature is zero though the metric $g_{+-}$
takes
an infinite value on this line. We obtain singularity if $x^+$ or $x^-$ takes
infinite value.
It is convenient to change variables: $x^{\pm}=(y^{\pm})^{-1}$. Then, the
metric and
curvature take the form
\begin{eqnarray}
&&ds^2={1 \over (y^+y^-)^2} [{\sqrt{am} \over 1 -{(1-c)am \over 4ey^+y^-}}]^{1
\over 1-c}
dy^+dy^- \nonumber \\
&&R=1/e [{1 \over \sqrt{am}} (1-{(1-c) am \over 4e y^+y^-})]^{2c-1 \over 1-c}
\end{eqnarray}
In the coordinates $(y^+,y^-)$ the singularity lies  on the light cone
$y^+y^-=0$. Asymptotically
(for $y^+y^- \rightarrow \infty$) this space-time is of constant curvature.

\bigskip

{\it The special case is $c=1/2$}.

One can see from (5.11) that $w^{\pm}(x^{\pm})$ transform as usual scalar
fields.
Hence, one cannot put $w=0$. Taking into account (5.13), (5.14) we obtain for
eq.(5.3):
\begin{eqnarray}
&&\{ A;x^+ \} +{1 \over 2} (\partial_{+} w)^2 +T^{\psi}_{++}-t_{+}=0
\nonumber \\
&&\{ B; x^- \} +{ 1\over 2} (\partial_{-}w)^2 +T^{\psi}_{--}-t_{-}=0
\end{eqnarray}
The metric and curvature take the form:
\begin{eqnarray}
&&ds^2={A'B' e^{w^+} e^{w_{-}} \over (1-{AB \over 8e})^2} dx^+dx^-
\nonumber \\
&&R=1/e e^{-w^+} e^{-w^-}
\end{eqnarray}
We may use the coordinate freedom to choose $A$ and $B$ as new coordinates:
$u=A(x^+), \ v=B(x^-)$. Since under $x^+ \rightarrow y^+(x^+)$ the
Schwarzian derivative transforms as follows [18]:
\begin{equation}
\{ A; x^+ \} \rightarrow (\partial_{+} y^+)^2 \{A; y^+ \} +\{y^+; x^+ \}
\end{equation}
eqs.(5.26) are rewritten in the following form:
\begin{eqnarray}
&&{ 1\over 2} (\partial_{u} w)^2 +T^{\psi}_{uu}=t_{u} \nonumber \\
&&{ 1\over 2} (\partial_{v} w)^2 +T^{\psi}_{uu}=t_{v}
\end{eqnarray}
Note that inhomogeneous piece of law (5.28) cancels in (5.26) with the
corresponding
transformation of $t_{\pm}$, so that finally we come to expression (5.29).
In the new coordinates we have
\begin{eqnarray}
&&ds^2={ e^{w^+ (u)} e^{w_{-}(v)} \over (1-{uv \over 8e})^2} dudv
\nonumber \\
&&R=1/e e^{-w^+(u)} e^{-w^-(v)}
\end{eqnarray}
In general, the solution of eqs.(5.29) depends on the choice of boundary
conditions,i.e., on appropriate functions $t_{u}, \ t_{v}$. These functions
mean the flow of the Hawking radiation due to the falling matter
with energy-momentum tensor $T^{\psi}_{\mu\nu}$. Physically, it seems to be
reasonable to assume that the energy back radiated cannot be larger than
the energy of the falling matter, i.e. $t_{\pm}  \leq T^{\psi}_{\pm\pm}$. From
(5.29) we obtain that unique possibility is the following: $T^{\psi}_{\pm\pm}=
t_{\pm}$. Hence, one gets $w_{\pm}=const$ and, consequently, the total
space-time
is of the constant curvature: $R=e^{-(w+1)}$. Notice, that only for $c=1/2$
there
exists a constant curvature (de Sitter) solution of equations (5.3)-(5.5).

\bigskip

{\it The other special case is $c=1$}.

Then, as one can see from (5.6) and (5.7) we obtain
\begin{equation}
\partial_{+}\partial_{-} (\sigma +\phi)=0
\end{equation}
This equation has the solution
\begin{equation}
\sigma+\phi=w=w^{+}(x^+)+w^{-}(x^-)
\end{equation}
Inserting this into (5.7) we get an equation on conformal factor $\sigma$:
\begin{equation}
\partial_{+}\partial_{-}\sigma ={1 \over 4e} e^w={1 \over 4e} e^{w^-} e^{w^+}
\end{equation}
which has the solution:
\begin{equation}
\sigma={ 1 \over 4e} \int_{}^{x^+}e^{w^+(z^+)}dz^+ \int_{}^{x^-}
e^{w^-(z^-)} dz^-    +\alpha (x^+)+\beta (x^-)
\end{equation}
We use the coordinate freedom to put $\alpha(x^+)=0,\ \beta(x^-)=0$.
One can see that $\partial_{\pm} \sigma$ satisfy the following equation
\begin{equation}
\partial^2_{\pm}\sigma=\partial_{\pm}w_{\pm} \partial_{\pm}  \sigma
\end{equation}
Taking this into account and putting (5.32), (5.34) into (5.3) we obtain that
$w_{\pm}$
satisfy the following equations:
\begin{equation}
\partial_{\pm}^2 w_{\pm}=-(T^\psi_{\pm\pm} -2t_{\pm})
\end{equation}
The general solution of (5.36) takes the form
\begin{equation}
w_{\pm} (x^\pm)=-\int_{}^{x^\pm}du \int_{}^{u}(T^\psi_{\pm\pm}-2t_{\pm})dz
\end{equation}
If matter doesn't contribute (i.e. the right-hand side of (5.36) is zero),
then
\begin{equation}
w^+=ax^+ +b , \ w^-= mx^- +d
\end{equation}
where $a,b,m,d$ are constants. Below we consider the case $a,m >0$.

Let us now consider the $\delta$-like matter contribution ($t_{\pm}$ are
putted to zero) $T^\psi_{++}= \lambda \delta(x^+-x^+_0)$, $T^\psi_{--}=0$
 ($\lambda>0$). Then, $w^-=mx^- +d$ for all $x^+$ and $w^+$ takes the form
\begin{eqnarray}
w^{+}&=& ax^+ +b, \ \ \ if \ \ \ x^+ < x^+_0 \nonumber \\
 &=& (a-\lambda) x^+ +b + \lambda x^+_0, \ \ \ if \ \ \ x^+>x^+_0
\end{eqnarray}
Choosing the integrating constants in (5.34) to be zero
we have correspondingly for $\sigma$:
\begin{eqnarray}
\sigma &=& {1 \over 4e am} e^{mx^- +d} e^{ax^++b}, \ \ \ if \ \ \ x^+<x^+_0
\nonumber \\
&=&{1 \over 4e m(a-\lambda)} e^{mx^- +d} [e^{(a-\lambda)x^+ +b+\lambda
x^+_0} -{\lambda \over a} e^{ax^+_0 +b}], \ \ \ if \ \ \ x^{+}>x^+_0
\end{eqnarray}
We see that the metric $g_{+-}={1 \over 2}e^\sigma$ is  everywhere positive and
regular for any finite $x^+, x^-$.

It is worth observing that the scalar curvature $R=1/e e^{w-\sigma}$ can be
written in the
form:
\begin{equation}
R={ 1 \over e} \chi e^{-\alpha \chi}
\end{equation}
where we introduced the function $\chi$ ($\chi>0$) taking  the form
\begin{eqnarray}
\chi &=& e^{(b+d)} e^{(mx^{-} +ax^+)}, \ \ if \ \ x^+ < x^+_0 \nonumber \\
 &=& e^{(b+d)} e^{(mx^- +(a- \lambda)x^+ + \lambda x^+_0)}, \ \ if \ \ x^+ >
x^+_0
\end{eqnarray}
and function $\alpha$ is $\alpha={1 \over 4e am}$ for $x^+<x^+_0$
and
\begin{equation}
\alpha={1 \over 4em (a-\lambda)}[1-{\lambda \over a}
e^{(\lambda-a)(x^+-x^+_0)}]
\end{equation}
for $x^+>x^+_0$.
One can see that $\alpha$ is positive both for $\lambda<a$ and $\lambda>a$.
Moreover it takes positive finite value in the limit $\lambda \rightarrow a$:
\begin{equation}
\alpha \rightarrow {1 \over 4em}[{1 \over a} +x^+-x^+_0], \ \ \ if \ \ \
\lambda \rightarrow a
\end{equation}
Hence, we obtain that the function $\alpha$ is positive for all $x^+$ and the
curvature
$R$ (5.41) is finite for all $x^+, x^-$. We obtain the asymptotically flat
space-time
which is free from singularity and horizons.

Thus, the solution of equations (5.3)-(5.5) for $c=1$ describes everywhere
regular space-time. This case gives us good example when the quantum
corrections
(taken into account in the form of the Polyakov-Liouville term in the action
(5.2))
can really remove the space-time singularity of the classical (black hole)
solution. It should be noted that this result essentially depends on the
quantum
state or, equivalently, on the choice of appropriate boundary conditions
(functions
$t_{\pm}$). In the case under consideration the choice was to get
asymptotically
flat space-time.

\bigskip

Some remarks are in order. As we have seen in Section 3. (eq.(3.1))
for action (1.1) the value $(1/f'')$ is effectively loop expansion parameter
for gravity.
The semiclassical approximation (5.2) is valid under the condition: $| 1/f''|
<<N$.
For $f=R \ln R$ we have $(f'')^{-1}=R$. Consequently, we obtain the  condition:
$|R|<<N$.
This condition is rather natural and it means that semiclassical
approximation works far from the points where the curvature infinitely grows.
It is seen from the above consideration that for fixed very large $N \ (c>>1)$
there necessary exists a region near the space-time singularity where this
condition
is not valid and hence the semiclassical approximation  failed. However, we can
see
that for $N=48$ $(c=1)$ the curvature $R$ (5.41) is bounded and has a maximum
value:
$R_{max}=(e^2 \alpha)^{-1}$. We have that $\alpha=(4eam)^{-1}$ where $(am)$ is
an integrating constant. Thus, we obtain that a semiclassical approximation is
valid
for $c=1$ everywhere in the space-time if $(am)<<12e$. The last condition can
always be held by an appropriate choice of the integrating constant $(am)$.
For $N=24 \ (c=1/2)$ the curvature $R$ was shown to be constant: $R=e^{-(w+1)}$
{}.
By an appropriate choice of the constant $w$ one can control the condition
$R<<N$,
so the semiclassical
approximation is correct also in this case.

\bigskip

\section{Discussion}
\setcounter{equation}0

In resume, we have obtained that the preliminary hope that one-loop quantum
corrections remove the classical black hole singularity is not
realized for very large $N \ (c>>1)$. The space-time singularity is still
present
in the general solution for the action (5.2) in this regime. However, something
interesting
happens when $N$ takes some finite (not very large) values. We have shown that
for $N=48$ $(c=1)$ the solution of the system (5.2) describes  geodesically
complete
space-time regular everywhere. The corresponding scalar curvature
(5.41) takes only finite values. In the other case, when $N=24$ ($c=1/2$)
the semiclassical action (5.2) describes the (de Sitter) space-time of constant
curvature.
This space-time is also obviously free from singularities. Remember that $N$ is
the number
of scalar fields or, more generally, $N$ is the number of sorts of particles
in a matter multiplet.

We conclude with some remarks in the order of discussion.
It seems to be reasonable to consider the requirement of space-time regularity
as some kind of principle: " {\it The space-time singularities must be absent
in the complete quantum theory of gravity and matter} ". Then, our
semiclassical
analysis can be interpreted as that this "regularity principle" is not valid
in general. But it leads to some restrictions on the particle contents
of the theory. In the case under consideration, it constraints the number
of matter fields $N$. There are some well known principles in modern
physics which bound the particles spectrum: for example, the requirement
of anomalies cancellation. Therefore, it would not be very surprising if the
black
hole physics gives us  one more. In this paper, we have considered the
two-dimensional
case. However, the same situation can take place in four dimensions [15].

Of course, our study is just semiclassical and cannot be considered as
a strict proof. The analysis in the framework of the complete quantum theory
is necessary. However, the above consideration seems to be an argument
in favor of the hypothesis on the relation between absence of the space-time
singularities
and particle spectrum of the theory.

\bigskip

\section{Acknowledgments}
I would like to thank Yu.Obukhov for comments on calculating
the Seeley coefficients for 4-th order differential operators.
I also thank L.Avdeev, D.Fursaev, D.Kazakov, M.Kalmykov, I.Volovich
for very useful discussions. This work was supported in part  by the grant
Ph1-0802-0920 of the International Science Foundation.

\newpage
\section{Figure Captions}

{\bf Fig.1}:  The shape of the function $\phi (R)$ in the vicinity of point
$R=R_0$ where $\phi'(R)=f''(R)=0$. There are two values $R_a, R_b$ of $R$ which
correspond to the same value $\phi$. So the inverse function $R(\phi)$ has two
branches.

{\bf Fig.2}:  The space-time near the time-like line $R=R_0$ ($\phi=0$) where
$\phi'(R)=f''(R)=0$. It consists of two sheets glued along the line $R=R_0$.

{\bf Fig.3}:  The space-time obtained by the falling of $\delta$-like impulse
of matter at $x^+=x^+_0$ on the black hole. For $x^+>x^+_0$ the singularity is
slightly deformed and asymptotically reaches the new horizon at
$x^-=x^-_{\infty}$.

{\bf Fig.4}:  The black hole space-time deformed by quantum backreaction for
$c>1$.
The singularity now is time-like and points of horizon satisfy the condition
$x^+x^-=0$. The regions I and III are asymptotically flat.

{\bf Fig.5}:  The space-time obtained by the falling of $\delta$-like impulse
of matter at $x^+=x^+_0$ on the black hole for $c>1$. The singularity for
$x^+>x^+_0$ is slightly shifted and asymptotically tends to new horizon at
$x^+=x^+_1$ and $x^-=x^-_{\infty}$.
\end{document}